\documentclass[twocolumn,showpacs,preprintnumbers,amsmath,amssymb]{revtex4}
\usepackage{graphicx}% Include figure files
\usepackage{dcolumn}% Align table columns on decimal point
\usepackage{bm}% bold math

\begin{document}

\newcommand{\kpar}{k_\parallel}

\title{Influence of Phonon dimensionality on Electron Energy Relaxation}% Force line breaks with \\

\author{J. T. Karvonen}
\author{I. J. Maasilta}
 \affiliation{Nanoscience Center, Department of Physics, P.O. Box 35, FIN-40014 University of Jyv\"{a}skyl\"{a}, Finland.}%Lines break automatically %%@
%or can be forced with \\


\begin{abstract}
We studied experimentally the role of phonon dimensionality on electron-phonon (e-p) interaction in thin copper wires evaporated either on suspended %%@
silicon nitride membranes or on bulk substrates, at sub-Kelvin temperatures. The power emitted from electrons to phonons was measured using sensitive %%@
normal metal-insulator-superconductor (NIS) tunnel junction thermometers. Membrane thicknesses ranging from  30 nm to 750 nm were used to clearly see %%@
the onset of the effects of two-dimensional (2D) phonon system. We observed for the first time that a 2D phonon spectrum clearly changes the %%@
temperature dependence and strength of the e-p scattering rate, with the interaction becoming stronger at the lowest temperatures below $\sim$ 0.5 K %%@
for the 30 nm membranes. 
\end{abstract}

\pacs{63.22.+m, 63.20.Kr, 85.85.+j}% PACS, the Physics and Astronomy
\maketitle
It is an established fact that at sub-Kelvin temperatures the thermal coupling between conduction electrons and the lattice becomes very weak %%@
\cite{gant}. This has significant implications for the operation of low-temperature detectors  and coolers \cite{giaz}, or for any solid-state systems %%@
where dissipation and cooling are relevant. Low-temperature electron-phonon (e-p) interaction has been  studied widely during the past decades, but %%@
mostly only for the case in which the phonons are fully three dimensional (3D) \cite{Roukes,wellstood,kanskar,schmidt}. However, due to significant %%@
advances in fabrication of thin suspended structures, many practical devices and detectors exist in which the phonons are expected to move freely only %%@
within the plane of a membrane, forming a quasi-2D system \cite{cleland}. The question how the two-dimensionality of the phonon modes influences e-p %%@
interaction has been addressed theoretically for certain cases \cite{belitz,johnson,glavin},  but no clear experimental observation of the effect has %%@
been reported to date, although several attempts have been made \cite{ditusa,kwon}.

In this paper, we show for the first time experimentally that the electron-phonon interaction clearly changes depending on the dimensionality of the %%@
phonons, as expected from theory.  E-p coupling was measured with the help of sensitive NIS tunnel junction thermometry \cite{Rowell}, for thin Cu %%@
wires on suspended  silicon nitride (SiN$_x$) membranes with thickness varying from 30 nm to 750 nm, which spans the transition from 2D to 3D phonons. %%@
In addition, samples with identical Cu wires on bulk substrates were also measured for comparison. For the thinnest membranes, the e-p interaction was %%@
{\em strengthened} in comparison with the bulk samples, and its temperature dependence changed significantly, as is predicted by the theory %%@
\cite{belitz,johnson,glavin}. The change was large enough to give indirect evidence that the dispersive ($\omega \sim k^2$), flexural modes of the %%@
membrane likely play a major role in the e-p interaction.

In the presence of stress-free boundaries, the bulk transversal
and longitudinal phonon modes (with sound velocities $c_t$ and $c_l$, respectively)
couple to each other and form a new set of eigenmodes, which in the case of a suspended
membrane are known as the horizontal shear modes ($h$), and symmetric ($s$) and antisymmetric
($a$) Lamb modes \cite{Auld}.
The frequencies $\omega$ for the $h$ modes are simply 
$\omega= c_t\sqrt{\kpar^2+(m\pi/d)^2}$,
where $\kpar$ is the wave vector component parallel to the membrane surfaces,
$d$ is the membrane thickness and the integer $m$ is the branch number. However, the dispersion
relations of the $s$ and $a$ Lamb modes cannot be given in a closed analytical form, but have to be 
calculated numerically. The lowest three branches, dominant for thin membranes at low temperatures, have low frequency analytical expressions: %%@
$\omega_h=c_t\kpar$, $\omega_s=c_s\kpar$, and $\omega_a=\frac{\hbar}{2m^\star}\kpar^2$, 
where $c_s=2c_t\sqrt{(c_l^2-c_t^2)/c_l^2}$ is the effective sound velocity of the $s$ mode, and
$m^\star=\hbar\left[2c_td\sqrt{(c_l^2-c_t^2)/3c_l^2}\right]^{-1}$ is an effective mass for the $a$-mode "particle". 
This lowest $a$-mode with its quadratic dispersion is mostly responsible for the non-trivial behavior of the e-p interaction \cite{johnson,glavin}. %%@
Note that already a single free surface affects the modes \cite{geller} and the e-p interaction \cite{qu}, as the bulk modes couple and form another %%@
new set of eigenstates, including the surface localized Rayleigh-mode. Thus, the widely observed result for e-p power flow $P=\Sigma V(T_e^5-T_p^5)$ %%@
from a metal volume $V$ with $T_e$ the electron and $T_p$ the phonon temperature, is not expected to hold even  for thin enough films on bulk %%@
substrates.

\begin{figure}[h]
\includegraphics[width=0.8\linewidth]{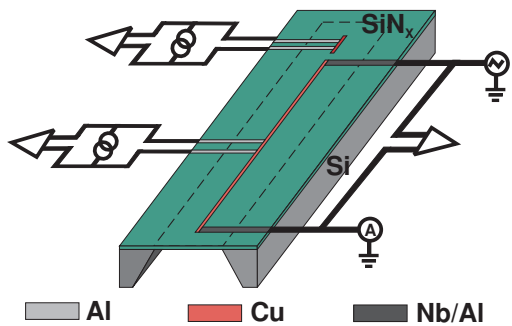}% Here is how to import EPS art
\caption{\label{fig:1}(Color online) A Schematic of the suspended samples and the measuring circuit. Red lines are the normal metal Cu, light gray Al %%@
for SINIS-junctions and dark gray Al or Nb for SN-junctions.}
\end{figure}

A schematic of the Cu wire samples on suspended silicon nitride membranes and the used measuring circuit is shown in Fig. \ref{fig:1}. 17 samples were %%@
made on either suspended membranes or bulk substrates, where nitridized (100) Si wafers with 30, 200 and 750 nm thick low-stress SiN$_{x}$ top layers  %%@
were used as the substrate for both cases. The suspension of the SiN$_x$ membranes (size $~$600$\times$300 $\mu$m$^2$) was achieved by  anisotropic %%@
backside wet etching of the silicon substate in KOH, and the metallic structures were fabricated using standard e-beam lithography and multi-angle %%@
shadow mask evaporation techniques. As the e-p interaction strength is sensitive to the thickness and disorder level of the metal \cite{JLTP}, we %%@
minimized its effect by evaporating the Cu wires of a specific thickness on all the different substrates simultaneously. Ultrathin Cu layers %%@
($t$=14-30 nm) were used to strengthen the effect of the thin membranes. The oxide layer forming the tunnel junction barriers was produced by thermal %%@
oxidation of Al. Table \ref{tab:table} presents the essential dimensions of the samples discussed in this paper, measured by scanning electron (SEM) %%@
and atomic force (AFM) microscopies. The electron mean free path $l$ was determined from the resistance of the wire at base temperature $~$ 60 mK, %%@
using the accurately measured dimensions of the wire.  

\begin{table}[h]
\caption{\label{tab:table}Parameters for samples. M= suspended SiN$_x$ membrane and B= bulk substrate. B6 had an oxidized Si substrate.}
\begin{ruledtabular}
\begin{tabular}{ccccccc}
Sample & SiN$_x$ $d$ & Cu $t$ & $V$  & $l$ & $\tau$(0.2K) & $\tau$(0.8K) \\
&(nm)&(nm)&[($\mu$m)$^3$]&(nm)&($\mu$s)& ($\mu$s)\\      
\hline
M1 & 30  & 14 & 2.71 & 5.7& 2.6 & 0.16\\
B1 & 30  & 14 & 2.46 & 4.9& 7.1 & 0.030\\
M2 & 200 & 14 & 2.44 & 4.6& 15.0& 0.11\\
B2 & 200 & 18 & 3.67 & 4.1& 6.4 & 0.045\\
\hline
M3 & 30  & 19 & 5.50& 11.2& 2.2& 0.30\\
B3 & 30  & 19 & 4.62&  9.8& 4.3& 0.034\\
M4 & 750 & 22 & 6.09& 10.3& 3.1& 0.030\\
B4 & 750 & 22 & 5.87&  8.7& 3.9& 0.013\\
\hline
M5 & 30 & 32 &  6.09&   22& 1.8& 0.31\\
B5 & 30 & 32 &  5.09&   19& 2.7& 0.038\\
B6 & -  & 32 &  7.10&   22& 1.6& 0.031\\
\end{tabular}
\end{ruledtabular}
\end{table}

 We used the hot-electron technique \cite{Roukes} to measure the e-p interaction by overheating the electrons by Joule heat power $P$ and measuring %%@
the resulting electron temperature $T_e$. All the samples had two electrically isolated Cu normal metal wires next to each other (Fig. 1). The longer %%@
wire ($L=500 \mu$m) was heated by applying a slowly ramping voltage across the pair of superconducting Nb (or Al) leads in direct metallic contact to %%@
Cu, forming SN junctions. These junctions provide excellent electrical, but very poor thermal conductance due to Andreev reflection, as the junctions %%@
are biased within the superconducting gap $\Delta$. Thus, due to the lack of outdiffusion of electrons and the long length of the wire, input heat is %%@
distributed uniformly in the interior of the wire and the electron gas cools dominantly by phonons, instead of diffusively \cite{Hoffmann} or by %%@
thermal photons \cite{meschke}. Since $L>>L_{e-e}$, the electron-electron scattering length, electron temperature is also well defined without %%@
complications from non-equilibrium \cite{Pothier}. In our sample geometry the electron temperature is measured with two additional Al leads forming a %%@
NIS tunnel junctions pair (SINIS) in the middle of heated wire, as a function of input Joule power $P=IV$ measured in a four probe configuration. The %%@
purpose of the short Cu wire, with additional SINIS thermometer on it, is to give an estimate of the local phonon temperature $T_p$, as the e-p power %%@
flow depends on both $T_e$ and $T_p$.

The current-biased Al SINIS thermometer is ideally suited to measure temperature below a few Kelvins,  \cite{giaz} due to its high sensitivity (in our %%@
DC measurement $\sim$ 0.1 mK at 0.1 K) and low power dissipation. In addition, for all the data here, the SINIS voltage vs. temperature response %%@
follows the BCS theory without fitting parameters very accurately at least down to $\sim$ 0.2 K, where typically saturation sets in. This saturation %%@
depends on the strength of the e-p interaction (size of thermometer and type of substrate) and the amount of filtering, and thus we conclude that it %%@
is most likely caused by external noise heating. For this reason we take the most conservative approach and assume that all saturation is caused by %%@
it, in which case we  can use BCS theory to convert the measured voltage data for all temperatures.

Even if the electrons lose their energy overwhelmingly to the phonons in our sample geometry, it is still possible that the measured temperature is %%@
not only determined by the e-p interaction. This is because the emitted phonons could be removed so ineffectively from the membrane that the phonon %%@
transmission becomes a bottleneck for the energy flow.  Bulk scattering of phonons at low temperatures is very weak \cite{cleland}, even for thin %%@
disordered membranes \cite{thomas}, as is boundary resistance
for thin films on bulk substrates \cite{swartz,noteb}. In contrast, almost nothing quantitative is known about the boundary resistance between a thin %%@
metal film and a thin 2D membrane, or between a thin 2D membrane and a bulk substrate. However, it seems clear that if the combined metal film and %%@
membrane thickness is below the thermal wavelength of the phonons, the phonon modes in the two materials are strongly coupled, leading to an %%@
effectively non-existent boundary resistance.  Hence, if we check that the membrane temperature $T_p$ is not too high compared to $T_e$ (effective %%@
enough hot phonon removal), we can be confident that the measured $T_e$ reflects the e-p interaction.       

\begin{figure}[h]
\includegraphics[width=0.99\linewidth]{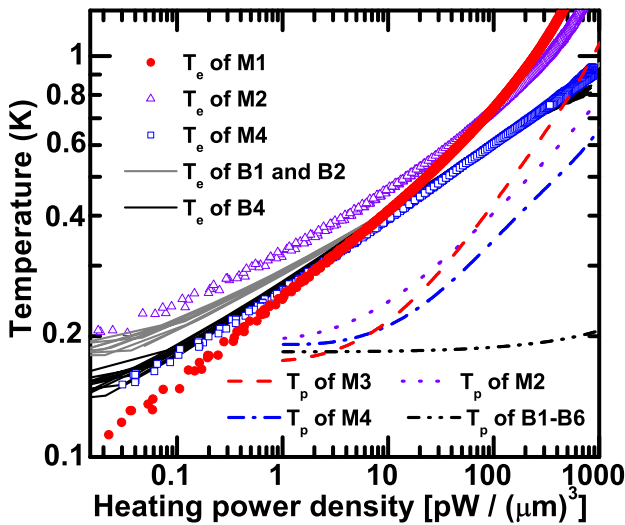}% Here is how to import EPS art
\caption{\label{kuva1} (Color online) Measured electron and phonon temperatures $T_e$ and $T_p$ versus the applied heating power density in %%@
log-log-scale.}
\end{figure}

Figure \ref{kuva1} shows the main result of the measurements, with $T_e$ and $T_p$ plotted vs. the heating power density $p=P/V$ for all membrane %%@
thicknesses (30 nm, 200 nm and 750 nm). In addition, data from a few representative bulk samples are shown. Compared to the corresponding bulk %%@
substrate sample (B4), $T_e$ of the 750 nm membrane (M4) shows no difference at all, and it effectively behaves as bulk. This is reasonable, because %%@
for the 750 nm membrane  the estimated dimensionality cross-over temperature  \cite{KuhnPhysRevB,NIMA} $T_{cr}=\hbar c_t/(2k_Bd)$ is $\sim 30$ mK, %%@
with $c_t=6200$ m/s for SiN. The phonon temperatures $T_p$, however, show a big difference: The bulk samples show almost no response from the %%@
saturation value of the thermometer $\sim$ 190 mK, whereas the membrane phonons heat up measurably, most likely due to the boundary resistance between %%@
the membrane and the bulk. Nevertheless, this increase in $T_p$ for all samples is small enough not to influence the e-p interaction. For the 200 nm %%@
thick membrane (M2) ($T_{cr} \sim$ 110 mK),   at low heating power densities [$p < 40$ pW/($\mu$m)$^3$] the temperature dependence follows the %%@
behavior of the bulk sample (B2), although with a difference in the absolute value. This shows  that the strength of the e-p coupling weakens compared %%@
to the bulk. At higher powers and temperatures ($p > 40$ pW/($\mu$m)$^3$, where $T_e > 0.6$ K), $T_e$ starts to increase more rapidly in the membrane %%@
sample, most likely due to the boundary resistance effects. The phonons in the 30 nm thick membrane sample (M1) are expected to be in the 2D limit at %%@
low temperatures ($T_{cr} \sim 0.5$K), and a clear sign of this can be seen in Fig. \ref{kuva1} as a strongly different behavior of the measured $T_e$ %%@
vs. $p$ curve with respect to all other samples. Below $\sim$ 6 pW/($\mu$m)$^3$ the e-p coupling is notably stronger ($T_e$ lower) than in the %%@
corresponding bulk (B1) or any other sample, but again at highest temperatures the influence of other effects starts to dominate over the e-p %%@
coupling.

\begin{figure}[ht!]
\includegraphics[width=0.9\linewidth]{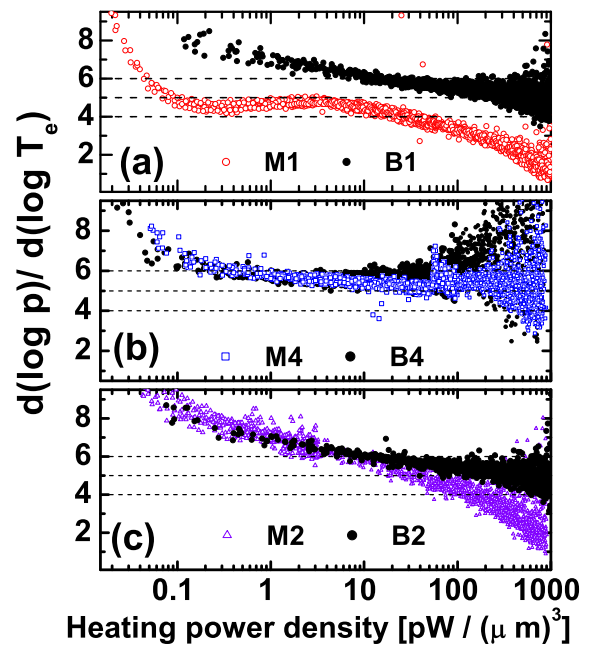}% Here is how to import EPS art
\caption{\label{kuva2} (Color online) Numerical logarithmic derivatives of the measured data in Fig. \ref{kuva1}. (a) $T_e$ data for M1 and B1,
  (b) $T_e$ data for M2 and B2, (c) $T_e$ data for M4 and B4.} 
\end{figure}

To study the temperature dependence of the data in Fig. \ref{kuva1} more accurately, we plot  the logarithmic derivatives $d(\log p)/d(\log T_e)$ in %%@
Fig. \ref{kuva2} (a)-(c). For low heating powers ($T_e^n >> T_p^n$)  $P_{e-p} \approx T_e^n$, where $n$ is the power law of the e-p interaction, thus %%@
in that regime $d(\log p)/d(\log T_e)=n$. Typically this exponent is $n \approx 5$ for thicker ($t > 30$ nm) metal films on bulk substrates %%@
\cite{Roukes, wellstood, JLTP}, if the disorder in the film is not too strong \cite{schmid,sergeev,taskinen}.  From Fig. \ref{kuva2} (a) we first of %%@
all see that for the 30 nm membrane sample M1, the difference to the bulk sample B1 is very clear. The M1 data has a plateau of $n\sim 4.5$ between %%@
$p=$ 0.1 - 6 pW/($\mu$m)$^3$, while for B1, $n$ continuously decreases from much higher values. Note that the strong increase of $d(\log p)/d(\log %%@
T_e)$ below $p \sim 0.1$ pW/($\mu$m)$^3$ is caused by the saturation of the $T_e$ measurement, and not by the e-p interaction. The point where $n$ %%@
starts deviating from $n=4.5$ corresponds to $T_e \approx$ 0.4 K, which is surprisingly consistent with the estimated $T_{cr} \sim$ 0.5 K.  In %%@
contrast, the temperature dependence of the 200 nm membrane (M2) and bulk (B2) samples [Fig. \ref{kuva2} (b)] are identical with each other and with %%@
the 30 nm bulk sample (B1), as long as the e-p interaction is dominant (up to 40 pW/($\mu$m)$^3$). The 750 nm membrane (M4) and bulk (B4) samples also %%@
give  identical values of $n$ [Fig. \ref{kuva2} (c)]. The difference between sample pairs M4,B4 and M2,B2 is caused by the Cu wire thickness, which is %%@
expected to influence the temperature dependence strongly \cite{qu,sergeev}.

\begin{figure}[h]
\includegraphics[width=0.99\linewidth]{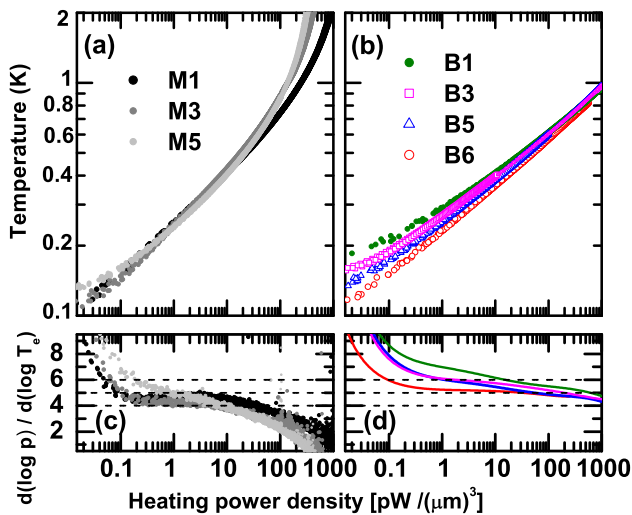}% Here is how to import EPS art
\caption{\label{kuva3} (Color online) (a) $T_e$ versus $p=P/V$ for 30 nm membrane samples M1,M3,M5. (b) $T_e$ versus $p$ for bulk samples, from top to %%@
bottom B1 (top), B3, B5 and B6 (bottom). (c) $d(\log p)/d(\log T)$ of the data in (a). (d) $d(\log p)/d(\log T)$ of the data in (b). From top to %%@
bottom: green line B1 (top), magenta B3, blue B5, Red B6 (bottom). In (d) noise has been filtered to help the eye.}
\end{figure}

Finally, we discuss the effect of the Cu wire thickness on the measured e-p interaction. The results for the thinnest 30 nm membrane samples, with Cu %%@
thickness $t=$ 14,19 and 32 nm are shown in Figs \ref{kuva3} (a) and (c). It is apparent that the metal film thickness has only a minor effect on the %%@
e-p interaction on thin membranes, and only influences the boundary resistance in the 3D limit, by increasing its effect for thicker $t$, as expected. %%@
However, for wires on bulk substrates, Figs \ref{kuva3} (b) and (d), the effect of the Cu wire thickness on e-p interaction is more profound. The %%@
thinner the Cu film, the more its temperature dependence deviates from $n=5$, which, for comparison, is observed for a more typical $t=32$ nm Cu wire %%@
on oxidized Si (B6). This behavior is qualitatively consistent with the predicted effect of the surface phonon modes \cite{qu}, but could also depend %%@
on the disorder, as the thickening of the film increases the mean free path $l$ (Table \ref{tab:table}) and pushes the sample closer to the clean %%@
limit. An apparent exponent as high as $\sim $ 7 could possibly be explained by the combination of strong disorder and surface modes, but again, %%@
detailed theory is lacking. 

In conclusion, we have obtained the first clear evidence that the electron-phonon interaction at low temperatures changes quite significantly when the %%@
phonon modes become two-dimensional. To quantify the effects, the electron thermal relaxation times $\tau=\gamma V T_e/(dP/dT_e)$, where $\gamma=100$ %%@
J/K$^2$m$^3$ for Cu, are presented in Table \ref{tab:table} for all the samples at two temperatures $T_e=0.2$ and 0.8 K. At $T_e <$ 0.5 K, the %%@
thinnest membranes can have a a factor 2-3 strengthening effect, whereas at higher temperatures the thermal relaxation from membranes can be an order %%@
of magnitude weaker compared to bulk samples. The membrane close to transition region ($d$=200 nm) was shown to have a weaker ($\sim$ factor of two) %%@
e-p interaction strength than the bulk samples. Thinning the metal film on bulk substrates also leads to a sizeable weakening of the e-p interaction. 
The observed power law exponent for the 2D limit is consistent with $n\approx$ 4.5, and is much smaller than the corresponding bulk exponent $n= %%@
6..7$. A reduction by more than a factor one gives indirect evidence of the importance of the flexural, dispersive Lamb-modes for the membrane %%@
electron-phonon interaction, in agreement with theory \cite{johnson, glavin}.

Discussions with T. K\"uhn and A. Sergeev and technical assistance by H. Niiranen are acknowledged. This work was supported by the Academy of Finland %%@
project Nos. 118665 and 118231, and by the Finnish Academy of Sciences and Letters (J.T.K.).
%\bibliography{physrevlett}% Produces the bibliography via BibTeX.

\pagebreak

\end{document}